\documentclass[twocolumn,trackchanges,twocolappendix]{aastex63}

\submitjournal{ApJ}
\usepackage{amsmath}
\usepackage{booktabs}
\usepackage{float}
\usepackage{longtable}
\usepackage{amssymb}

\begin{document}
\title{Accretion-induced Collapse of Dark Matter-admixed Rotating White Dwarfs: Dynamics and Gravitational-wave Signals}

\author[0000-0003-2776-082X]{Ho-Sang Chan}
\affiliation{Department of Physics and Institute of Theoretical Physics, The Chinese University of Hong Kong, Shatin, N.T., Hong Kong S.A.R.}
\author[0000-0002-1971-0403]{Ming-chung Chu}
\affiliation{Department of Physics and Institute of Theoretical Physics, The Chinese University of Hong Kong, Shatin, N.T., Hong Kong S.A.R.}
\author[0000-0002-4972-3803]{Shing-Chi Leung}
\affiliation{Department of Mathematics and Physics, SUNY Polytechnic Institute, 100 Seymour Road, Utica, New York 13502, USA}
\affiliation{TAPIR, Mailcode 350-17, California Institute of Technology, Pasadena, CA 91125, USA}

\begin{abstract}
We present two-dimensional hydrodynamic simulations of the accretion-induced collapse (AIC) of rotating white dwarfs admixed with an extended component of dark matter (DM) comprising of sub-GeV degenerate fermionic DM particles. We find that the DM component would follow the collapse of the normal matter (NM) component to become a bound DM core. Thus, we demonstrate how a DM-admixed neutron star could form through DM-admixed AIC (DMAIC) for the first time, with the dynamics of DM taken into account. The gravitational-wave (GW) signature from the DMAIC shows distinctive features. In the diffusive DM limit, the DM admixture indirectly suppresses the post-bounce spectral peak of the NM GWs. In the compact DM limit, the collapse dynamics of the DM in a Milky Way event generate GWs that are strong enough to be detectable by Advanced LIGO as continuous low-frequency ($< 1000$ Hz) signals after the NM core bounce. Our study not only is the first-ever computation of GW from a collapsing DM object but also provides the key features to identify DM in AIC events through future GW detections.
\end{abstract}



\section{Introduction} \label{sec:intro}


\subsection{Dark Matter-admixed Astrophysical Objects} \label{subsec:darkstar}

It is widely believed that dark matter (DM) constitutes the major mass-energy component of galaxy clusters \citep{2006ApJ...648L.109} and large-scale structures of the Universe \citep{1985ApJ...292..371D}. Besides terrestrial experiments, physicists are tackling the DM problem through astrophysical observations. It is shown that in a region with a high concentration of DM particles, DM could be captured by normal matter \citep[NM,][]{doi:10.1063/1.4868744, 2019Ap&SS.364...24A}. Therefore, it is natural to expect stellar objects composed of NM and DM. There have been extensive theoretical studies on the possible effects of the DM admixture on the stellar evolution \citep{2019ApJ...879...50L, 2020arXiv200904302S, 2021MNRAS.tmp..854R}. Unusual stellar objects consistent with these models might hint at the existence of DM-admixed stars. Furthermore, there are studies utilizing DM-admixed star models to understand the properties of DM. For instance, \citet{PhysRevD.105.123010} proposed a method for inferring the DM particle mass by measuring the tidal deformability of neutron stars. Using the DM-admixed neutron star model, \citet{2013PhRvD..87e5012B} and \citet{2013PhRvD..87l3507B} gave constraints on the bosonic DM particle mass and annihilation cross section. These examples show that DM-admixed stellar objects could be a promising channel to probe astrophysical DM.


\subsection{Rotating White Dwarfs} \label{subsec:rotationwd}

The majority of studies on white dwarfs (WD) assume they are not rotating, but observational evidence shows the opposite \citep{1998A&A...333..603S, 2003astro.ph..1539K}. It was suggested that WDs gain angular momentum through accretion from a companion star \citep{2003astro.ph..2232L, Yoon_2004} or mergers between two or more WDs \citep{2019Natur.569..684G, 2020MNRAS.499L..21P}. Therefore, rotation is an important ingredient of the full picture of WD structure and evolution \citep{Yoon_2004, 2005A&A...435..967Y}. In addition, rotating WDs have been proposed to be progenitors of super-luminous thermonuclear supernovae because rotating WDs could support more mass than their traditional Chandrasekhar limit \citep{2010A&A...509A..75P, 2014MNRAS.445.2340W, 2018A&A...618A.124F}. Recently, the effects of the strong magnetic field on the equilibrium structures of WDs have been studied \citep{PhysRevD.92.083006, 2016MNRAS.456.3375B, 10.1093/mnras/stx781}, for which the WD rotation takes a critical role.


\subsection{Accretion-induced Collapse} \label{subsec:aic}

It was widely believed that a WD would undergo a thermouclear explosion when its mass is approaching the canonical Chandrasekhar limit. However, if the WD contains an oxygen-neon (O-Ne) core, Accretion-Induced Collapse (AIC) is possible as its mass increases towards the Chandrasekhar limit through stable accretion from a companion object \citep{1991ApJ...367L..19N, 2018MNRAS.481..439W, 2019MNRAS.484..698R}, though a binary WD merger seems to be another possible scenario \citep{2020MNRAS.494.3422L}. The collapse is triggered by electron capture in the degenerate matter, weakening the electron degenerate pressure \citep{2017ApJ...843..151B}. On the other hand, pycnonuclear burning is also possible in such an extremely dense core. Hence the ultimate fate of an O-Ne WD would depend on the competition between nuclear runaway and electron capture \citep{2020RAA....20..135W}. However, it was later found that the central temperature of O-Ne WDs is insufficient for explosive O-Ne burning \citep{2018RAA....18...36W}. Even if deflagration occurs, it fails to unbind the WD, which directly leads to a collapse for a wide range of parameters \citep{Leung2019ECSN, Zha2019, Leung2020ECSN}. \\

Besides the iron-core collapse of massive stars, the AIC of WDs has been proposed as another channel for forming neutron stars. However, AIC is much less luminous than typical core-collapse supernovae. The small amount of nickel synthesized indicates that AICs are usually faint transients \citep{2010MNRAS.409..846D}. On the other hand, AIC emits radio signatures \citep{2016ApJ...830L..38M} and has been hypothesized as a source candidate for Fast Radio Bursts \citep{2019ApJ...886..110M} and Millisecond Pulsars \citep{2022MNRAS.510.6011W}. Electromagnetic-wave detection of AIC would be a challenging but possible task. One possible way to search for AIC is by neutrino detection because a neutrino burst should accompany AIC after the WD dynamical collapse \citep{1992ApJ...388..164D}. The burst luminosity could be as large as $10^{55}$ erg s$^{-1}$ \citep{alma991040185904203407, 2006ApJ...644.1063D}. On the other hand, the collapse dynamics of the compact iron core are expected to produce strong GW signals \citep{2005ApJ...625L.119O, 2009CQGra..26f3001O}. There have been some efforts to predict the GW signature from an AIC. \citet{2006ApJ...644.1063D} simulated 2D AIC with neutrino transport and estimated the GW emission from AIC via the Newtonian quadrupole formula. They concluded that LIGO-class detectors could detect Milky Way AIC events. \citet{2010PhRvD..81d4012A} found that the GW signals from an AIC show a generic ``Type III'' shape, though detailed neutrino physics have been omitted. 


\subsection{Motivations} \label{subsec:motivation}

Although DM-admixed neutron stars have been studied and applied to explain anomalous compact objects, there is still no in-depth research on their formation channel. Even though \citet{2019ApJ...884....9L, 2019ApJ...883...13Z} numerically investigated DM-admixed AIC (DMAIC), they assumed the DM component to be spherically symmetric and non-moving. As pointed out by \citet{2019ApJ...884....9L}, the stationary DM approximation may break down if the dynamical time scales for DM and NM become comparable, and the dynamical modelling of the DM becomes important. They also pointed out that there is a moment during the collapse in which the NM has a mass density comparable with that of the DM. Also, \citet{Chan_2021} showed that fermionic DM with a sub-GeV particle mass would produce a massive and extended component comparable in size to that of the NM. In such a scenario, modelling the DM dynamics would be necessary. In this study, we extend the multidimensional simulations by \citet{2019ApJ...883...13Z} to include also the dynamical evolution of the DM component. Our study aims to investigate if DMAIC could make a DM-admixed neutron star, with the DM motion taken into account, and to predict the corresponding GW signature to facilitate the search for DM through observing AIC in the future. The structure of the paper is as follows: we present the method of obtaining the progenitor and the tools for simulation in Section \ref{sec:method}. We then present the results of collapse dynamics and gravitational-wave signals in Section \ref{sec:results}. Finally, we conclude our study in Section \ref{sec:conclusion}.


\section{Methodology} \label{sec:method}


\subsection{Equations of Hydrostatic Equilibrium} \label{subsec:hydrostatic}

We compute DM-admixed rotating WDs (DMRWDs) as DMAIC progenitors by solving the Newtonian hydrostatic equations, including the centripetal force:
\begin{equation}
\begin{aligned}
    \vec{\nabla} P_{1} = - \rho_{1}\vec{\nabla}\Phi, \\
    \vec{\nabla} P_{2} = - \rho_{2}\vec{\nabla}\Phi + [\rho_{2}\omega_{2}(s)^{2}s]\hat{s}, \\
    \nabla^{2} \Phi = 4\pi G(\rho_{1} + \rho_{2}).
\end{aligned}
\end{equation}
Here, the subscript $i = 1(2)$ denotes the DM (NM) quantities, and $\rho$, $P$, $\omega$, and $\Phi$ are the density, pressure, angular speed, and gravitational potential of the fluid element. $s$ is the perpendicular distance from the rotation axis, and $\hat{s}$ is the unit vector orthogonal to and pointing away from that axis. The angular velocity is assumed to be a function of $s$ only. We consider the Newtonian framework because the rotation speed and compactness of WDs are low. \\

We follow \citet{1985A&A...146..260E}, \citet{1986ApJS...61..479H} and \citet{1994A&A...290..674A} to integrate the equations of equilibrium:
\begin{equation}
\begin{aligned}
    H_{i} + \Phi + \delta_{i2}h_{i}^{2}\psi_{i} = C_{i}, \\
    \int \frac{dP_{i}}{\rho_{i}} = H_{i}, \\
    \int \omega(s)_{i}^{2}s ds = -h_{i}^{2}\psi_{i}, \\
\end{aligned}
\end{equation}
where $C_{i}$ is an integration constant, $H$ is the enthalpy, $\psi$ is the rotational potential, and $h^{2}$ is a constant to be determined \citep{1986ApJS...61..479H}. We solve the equilibrium equations for the DM and NM using a two-fluid, self-consistent field method \citep{Chan_2022}.


\subsection{Rotation Rules} \label{subsec:rotation}

We have considered rotation profiles for the NM from \citet{1986ApJS...61..479H} and \citet{Yoshida_2019} including (1) the rigid rotation:
\begin{equation}
	\omega(s)_{2}^{2} = \Omega_{2}^{2},
\end{equation}
and (2) the ``Kepler'' profile:
\begin{equation}\label{eqn:keplerequation}
	\omega(s)_{2}^{2} \propto 1/(d^{3/2} + s^{3/2})^{2},
\end{equation}
which resembles a rapidly rotating core surrounded by an envelope rotating at its Keplerian limit. Here, $d$ is the rotating core radius. \\

We integrate the angular velocity to obtain the effective potential of the rigid rotation:
\begin{equation}
  \psi_{2} =  s^{2}/2.
\end{equation}
The effective potential for the Kepler rule is:
\begin{equation}
\begin{aligned}
\psi_{2} = -\frac{1}{9} \left[-\frac{6\sqrt{s}}{s^{\frac{3}{2}} + d^{\frac{3}{2}}} +  \frac{1}{d}\text{ln}\left(\frac{(\sqrt{d} + \sqrt{s})^{2}}{d + s -\sqrt{sd}}\right)\right. \\
\left.-\frac{2\sqrt{3}}{d}\text{tan}^{-1}\left(\frac{1-2\sqrt{s/d}}{\sqrt{3}}\right)\right].
\end{aligned}
\end{equation}


\subsection{Hydrodynamic Evolution} \label{subsec:hydro}

To simulate DMAIC, we solve the two-dimensional Euler equations assuming axial symmetry:
\begin{equation} \label{eqn:hydro}
\begin{aligned}
    \partial_{t}\rho_{i} + \nabla\cdot(\alpha \rho_{i} \vec{v_{i}}) = 0, \\
    \partial_{t}(\rho_{i}\vec{v_{i}}) + \nabla\cdot[\alpha \rho_{i} (\vec{v_{i}}\otimes\vec{v_{i}})] + \nabla(\alpha P_{i}) = \\ 
    -\alpha(\rho_{i} - P_{i})\nabla\Phi.
\end{aligned}
\end{equation}
Here, $\alpha = \text{exp}(-\phi/c^{2})$ is the lapse function with $c$ being the speed of light. It is used to mimic general relativistic time-dilation effects and has been applied to study the first-order quantum chromodynamics phase transition in core-collapse supernovae \citep{2020PhRvL.125e1102Z}. We also solve the advection equation for the NM total internal energy density $\tau_{2} = \rho_{2}\epsilon_{2} + \rho_{2} v_{2}^{2}/2$ and electron fraction $Y_{e}$:
\begin{equation}
\begin{aligned}
    \partial_{t}\tau_{2} + \nabla\cdot[\alpha(\tau_{2} + P_{2})\vec{v_{2}}] = -\alpha\rho_{2}\vec{v_{2}}\cdot\nabla\Phi, \\
    \partial_{t}(\rho_{2} Y_{e}) + \nabla\cdot(\alpha\rho_{2} Y_{e} \vec{v_{2}}) = 0.
\end{aligned}
\end{equation}
The gravitational potential $\Phi$ is solved by a multipole solver, for which we adopt the one by \citet{Couch_2013} that can reduce error by computing the potential at the cell center while using the mass density at that point. To mimic general relativistic strong-field effects, we use the modified case A potential \citep{2008A&A...489..301M} as an additional correction to the Newtonian potential:
\begin{equation}
\begin{aligned}
    \Phi \rightarrow \Phi - \langle\Phi\rangle + \Phi_{\text{TOV}, 1} + \Phi_{\text{TOV}, 2}, \\
    \langle\Phi\rangle = -4\pi \int_{0}^{\infty}dr'r'^{2}\frac{\langle\rho_{1} + \rho_{2}\rangle}{|r - r'|}.
\end{aligned}
\end{equation}
Here, $r$ is the radial distance and $\langle\rho_{1} + \rho_{2}\rangle$ represents the angular average of the total density. $\Phi_{\text{TOV}, i}$ for $i = $1, 2 are the relativistic corrections:
\begin{equation}
\begin{aligned}
    \Phi_{\text{TOV},i} = -4\pi \int_{0}^{\infty}\frac{dr'}{r'^{2}}\frac{1}{\Gamma_{i}^{2}} \left(\frac{m_{\rm TOV, i}}{4\pi} + r'^{3}P_{i}\right) \\ \left(1 + \epsilon_{i} + \frac{P_{i}}{\rho_{i}}\right), \\
    m_{\rm TOV, i} = 4\pi\int_{0}^{r}dr'r'^{2}\Gamma_{i}\rho_{i}(1 + \epsilon_{i}), \\
    \Gamma_{i} = \sqrt{1 + v_{r, i}^{2} - \frac{2m_{\rm TOV, i}}{r}},
\end{aligned}
\end{equation}
where $v_{r, i}$ is the radial velocity. We adopt a finite-volume approach to solve the hydrodynamic equation in spherical coordinates \citep{article}. We use the piece-wise parabolic method \citep{1984JCoPh..54..174C} to reconstruct primitive variables at the cell interface and the HLLE Riemann solver \citep{inbook} to compute fluxes across cell boundaries. The reconstruction and flux evaluation are done on a dimension-by-dimension basis. We discretize the temporal evolution using the method of lines where the strong stability-preserving 5-step, 4th-order Runge-Kutta method is implemented \citep{doi:10.1142/7498}. In addition to the (modified) Euler equation, we also append the internal energy equation for the NM:
\begin{equation}
\begin{aligned}
    \partial_{t}(\rho_{2}\epsilon_{2}) + \nabla\cdot\{\alpha[(\rho_{2}\epsilon_{2}) + P_{2}]\vec{v_{2}}\} = \\
    \vec{v_{2}}\cdot\nabla(\alpha P_{2}) - \alpha P_{2} (\vec{v_{2}}\cdot\nabla\Phi).
\end{aligned}
\end{equation}
It not only allows one to interpolate the internal energy density $\epsilon$ to the cell interface so that computational cost could be reduced but also reduces the error of $\epsilon$ due to advection \citep{2020JPhCS1623a2021Z}. We adopt a computational grid similar to that in \citet{2016ApJ...831...81S}, in which an analytic function describes the positions of the radial cell interfaces as:
\begin{equation}
    r_{i} = A_{t}\text{sinh}(x_{t}i/A_{t}).
\end{equation}
Here, $i$ is the cell index. We set $x_{t} = 0.5$ (in code unit)\footnote{One code unit in length equals to $1.4766839$ km} and $A_{t} = 150$ so that a central resolution of around $0.74$ km is provided, while a total of $500$ computational grids are used to contain the progenitor. We use $20$ grids to resolve the polar direction, which we find sufficient for ensuring convergence in GW signals for both the NM and DM. \\


\subsection{Micro-physics} \label{subsec:micro}

After mapping the density profiles of the NM and DM components computed from Section \ref{subsec:hydrostatic}, we assign an initial temperature profile to the NM \citep{2006ApJ...644.1063D}:
\begin{equation}
    T(\rho) = T_{c}(\rho/\rho_{c})^{0.35}.
\end{equation}
Here, $T_{c} = 10^{10}$ K and $\rho_{c} = 5\times10^{10}$ gcm$^{-3}$ are the central temperature and density, respectively. The core electron capture process initiates the AIC. We implement the parameterized electron capture scheme described in \citet{Liebendorfer_2005} to simulate such a process. In their work, $Y_{e}$ depends on $\rho_{2}$ as:
\begin{equation} \label{eqn:ecap}
\begin{aligned}
    x(\rho_{2}) = \text{max}\left[-1, \text{min}\left(1, \frac{2-\text{log}\rho_{2} - \text{log}\rho_{\alpha} - \text{log}\rho_{\beta}}{\text{log}\rho_{\alpha} - \text{log}\rho_{\beta}}\right)\right], \\
    Y_{e}(x) = \frac{1}{2}(Y_{b} + Y_{a}) + \frac{x}{2}(Y_{b} - Y_{a}) + \\
    Y_{c}[1 - |x| + 4|x|(|x| - 1/2)(|x| - 1)].
\end{aligned}
\end{equation}
Here, $\text{log}\rho_{\alpha}$, $\text{log}\rho_{\beta}$, $Y_{a}$, $Y_{b}$, and $Y_{c}$ are fitting parameters and are obtained by \citet{2019ApJ...884....9L} and \citet{2019ApJ...883...13Z}. We first assign an initial equilibrium $Y_{e}$ profile to the NM using Equation \ref{eqn:ecap}. We then start the electron capture process by updating $Y_{e}$ at each time step using the same equation. We force $Y_{e}$ to strictly decrease with time. We terminate the electron capture process once the core bounce condition \citep{Liebendorfer_2005} is achieved, which is when the core NM entropy is larger than $3k_{B}$, where $k_{B}$ is the Boltzman constant.


\subsection{Gravitational-wave Signals} \label{subsec:gwave}

We use the quadrupole formula in the weak-field approximation to compute the GW strain \citep{1990ApJ...351..588F, 1991A&A...246..417M}:
\begin{equation}
    h_{+} = \frac{3}{2}\frac{G}{Dc^{4}}\text{sin}^{2}\theta\frac{d^{2}}{dt^{2}}I_{zz}.
\end{equation}
Here, $D = 10$ kpc is the assumed distance and $\theta$ is the orientation angle of the collapsing DMRWD, and $I_{zz}$ is the moment of inertia tensor:
\begin{equation} \label{eqn:gwave}
    I_{zz} = \frac{1}{3}\int_{\text{All Space}}(\rho_{1} + \rho_{2})r^{2}P_{2}(\text{cos}\theta) d\tau.
\end{equation}


\subsection{Equations of State} \label{subsec:eos}

To simulate AICs, we first use the ideal degenerate Fermi gas EOS for equilibrium structure construction. Following the subsequent collapse dynamics, we use the nuclear matter EOS given by \citet{2011ApJS..197...20S}, widely used in simulating core-collapse supernovae and neutron star dynamics. We adopt the ideal degenerate Fermi gas EOS for the DM component \citep{PhysRevD.74.063003}. 


\section{Results And Discussion} \label{sec:results}

We define $\bar{t}$ as the time after the NM core bounce, and we terminate our simulations at $\bar{t} = 0.1$ s.


\subsection{The Diffusive Dark Matter Limit} \label{subsec:diffusive}

We have computed a series of DMRWD models as DMAIC progenitors. The stellar parameters of these progenitors have been listed in Table \ref{tab:tabmodels} for reference. The progenitors have DM mass fractions $\epsilon_{\text{DM}} = M_{\rm DM}/(M_{\rm DM} + M_{\rm NM})$, where $M_{\rm NM}$ ($M_{\rm DM}$) is the NM (DM) mass, from $0.01$ to $0.2$ and include rigidly-rotating and differentially-rotating DMRWDs with different $d$ as described in Equation \ref{eqn:keplerequation}. In particular, $d$ is chosen so that $\rho_{2} (r = d, \theta = \frac{\pi}{2}) = \alpha_{d}\rho_{2c}$. We choose $\alpha_{d} = 0.1$ and $0.01$. Another free parameter to be specified for these progenitors is the central angular velocity $\Omega_{c}$. We adjust this value for rigidly-rotating DMRWDs so that the corresponding pure NM progenitor almost rotates at the Keplerian limit and that a total mass of $\approx 1.8$ $M_{\odot}$ is achieved for a pure NM, differentially rotating WD. We fix the DM particle mass to be $0.1$ GeV for all of these progenitors. As shown in \citet{Chan_2021}, the fluid component formed by DM particles with such a mass will be more diffusive and comparable in size to that of the NM.


\subsubsection{The Collapse Dynamics} \label{subsubsec:simaic}

\begin{deluxetable*}{ccccccccccccc}[ht!]
\tablecaption{Stellar parameters for different DMAIC progenitors. They include rigid (labelled Rigid) and differentially (labelled Kepler) rotating DMRWDs. All progenitors have NM central density of $5\times 10^{10}$ gcm$^{-3}$. The DM particle mass is $0.1$ GeV. \label{tab:tabmodels}}
\tablewidth{0pt}
\tablehead{
\colhead{Model} & \colhead{$M_{\rm NM}$} & \colhead{$M_{\rm DM}$} & \colhead{$\alpha_{d}$} & \colhead{log$_{10}\rho_{1c}$} & \colhead{$\Omega_{c}$} &
\colhead{$R_{\rm eNM}$} & \colhead{$R_{\rm eDM}$} & \colhead{$\epsilon_{\rm DM}$} & \colhead{$t_{b}$} & \colhead{log$_{10}\rho_{1b}$} & \colhead{log$_{10}\rho_{2b}$} & \colhead{$M_{\rm PNS}$} \\
\colhead{-} & \colhead{($M_{\odot}$)} & \colhead{($M_{\odot}$)} & \colhead{-} & \colhead{(gcm$^{-3}$)} & \colhead{(s$^{-1}$)} &
\colhead{(km)} & \colhead{(km)} & \colhead{-} & \colhead{(ms)} & \colhead{(gcm$^{-3}$)} & \colhead{(gcm$^{-3}$)} & \colhead{($M_{\odot}$)}
}
\startdata
Rigid-NM & 1.477 & 0.000 & - & - & 10.8 & 1105 & - & 0 & 53.151 & - & 14.318 & 1.217 \\
Rigid-0.01 & 1.447 & 0.015 & - & 8.816 & 10.8 & 1098 & 400 & 0.01 & 53.424 & 11.078 & 14.317 & 1.194 \\
Rigid-0.03 & 1.416 & 0.044 & - & 8.980 & 10.8 & 1027 & 568 & 0.03 & 53.717 & 11.093 & 14.315 & 1.170 \\
Rigid-0.05 & 1.397 & 0.074 & - & 9.046 & 10.8 & 980 & 695 & 0.05 & 53.902 & 11.097 & 14.316 & 1.157 \\
Rigid-0.07 & 1.384 & 0.104 & - & 9.086 & 10.8 & 948 & 807 & 0.07 & 54.035 & 11.100 & 14.315 & 1.146 \\
Rigid-0.09 & 1.374 & 0.136 & - & 9.114 & 10.8 & 923 & 904 & 0.09 & 54.139 & 11.102 & 14.315 & 1.139 \\
Rigid-0.1 & 1.307 & 0.152 & - & 9.125 & 10.8 & 910 & 954 & 0.1 & 54.183 & 11.103 & 14.315 & 1.136 \\
Rigid-0.2 & 1.343 & 0.336 & - & 9.193 & 10.8 & 857 & 1379 & 0.2 & 54.496 & 11.107 & 14.313 & 1.119 \\
Kepler-NM-d001 & 1.770 & 0.000 & 0.01 & - & 32.5 & 1826 & - & 0 & 35.124 & - & 14.355 & 1.597 \\
Kepler-0.01-d001 & 1.725 & 0.017 & 0.01 & 8.853 & 32.5 & 1766 & 420 & 0.01 & 35.352 & 11.073 & 14.352 & 1.553 \\
Kepler-0.03-d001 & 1.662 & 0.051 & 0.01 & 9.016 & 32.5 & 1494 & 587 & 0.03 & 35.647 & 11.086 & 14.351 & 1.498 \\
Kepler-0.05-d001 & 1.623 & 0.085 & 0.01 & 9.082 & 32.5 & 1343 & 710 & 0.05 & 35.835 & 11.091 & 14.352 & 1.463 \\
Kepler-0.07-d001 & 1.596 & 0.120 & 0.01 & 9.122 & 32.5 & 1256 & 812 & 0.07 & 35.969 & 11.094 & 14.350 & 1.439 \\
Kepler-0.09-d001 & 1.577 & 0.156 & 0.01 & 9.150 & 32.5 & 1190 & 904 & 0.09 & 36.072 & 11.096 & 14.350 & 1.419 \\
Kepler-0.1-d001 & 1.569 & 0.174 & 0.01 & 9.162 & 32.5 & 1159 & 948 & 0.1 & 36.116 & 11.097 & 14.350 & 1.411 \\
Kepler-0.2-d001 & 1.518 & 0.379 & 0.01 & 9.232 & 32.5 & 1020 & 1352 & 0.2 & 36.424 & 11.101 & 14.349 & 1.362 \\
Kepler-NM-d001 & 1.771 & 0.000 & 0.1 & - & 45.2 & 1106 & - & 0 & 32.311 & - & 14.354 & 1.598 \\
Kepler-0.01-d01 & 1.727 & 0.017 & 0.1 & 8.860 & 45.2 & 1098 & 417 & 0.01 & 32.555 & 11.070 & 14.353 & 1.555 \\
Kepler-0.03-d01 & 1.677 & 0.052 & 0.1 & 9.026 & 45.2 & 1062 & 579 & 0.03 & 32.788 & 11.084 & 14.354 & 1.511 \\
Kepler-0.05-d01 & 1.647 & 0.087 & 0.1 & 9.094 & 45.2 & 1034 & 700 & 0.05 & 32.924 & 11.088 & 14.351 & 1.483 \\
Kepler-0.07-d01 & 1.625 & 0.122 & 0.1 & 9.135 & 45.2 & 1007 & 801 & 0.07 & 32.024 & 11.092 & 14.351 & 1.460 \\
Kepler-0.09-d01 & 1.609 & 0.159 & 0.1 & 9.164 & 45.2 & 987 & 892 & 0.09 & 33.101 & 11.095 & 14.352 & 1.446 \\
Kepler-0.1-d01 & 1.602 & 0.178 & 0.1 & 9.176 & 45.2 & 980 & 941 & 0.1 & 33.133 & 11.096 & 14.352 & 1.439 \\
Kepler-0.2-d01 & 1.556 & 0.389 & 0.1 & 9.247 & 45.2 & 916 & 1343 & 0.2 & 33.364 & 11.101 & 14.355 & 1.396
\enddata
\tablecomments{In this table, $R_{\rm eNM}$ ($R_{\rm eDM}$) is the equatorial radius of the progenitor for the NM (DM) component. $\rho_{\rm 1c}$ is the DM central density, $\epsilon_{\rm DM}$ is the DM fraction, and $t_{\rm b}$ is the bounce time. $\rho_{\rm 2b}$ ($\rho_{\rm 1b}$) is the maximum NM (DM) density at the core bounce. $M_{\rm PNS}$ is the proto-neutron star mass, defined as summing all the NM mass with $\rho_{2} > 10^{11}$ gcm$^{-3}$ at the end of the simulation.}
\end{deluxetable*}

\begin{figure}[ht!]
	\centering
	\includegraphics[width=1.0\linewidth]{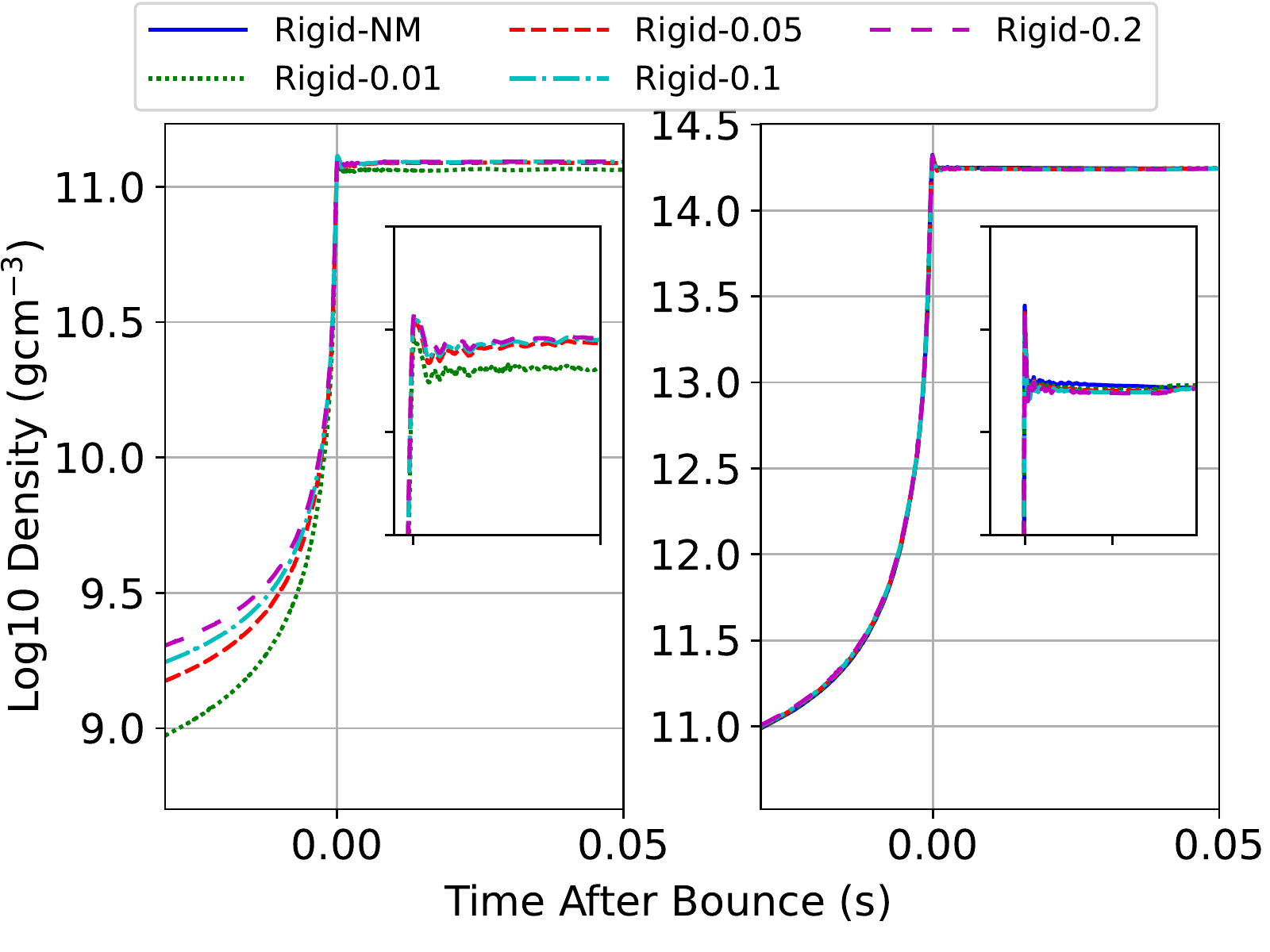}
	\caption{Evolution of the maximum density of the rigidly-rotating DMAIC models. The left (right) panel is for the DM (NM) component. Since there are only minimal deviations among different DM-admixed models, we show a magnified density evolution plot in each panel. \label{fig:nmrhocentralrigid}}
\end{figure}

\begin{figure}[ht!]
	\centering
	\includegraphics[width=1.0\linewidth]{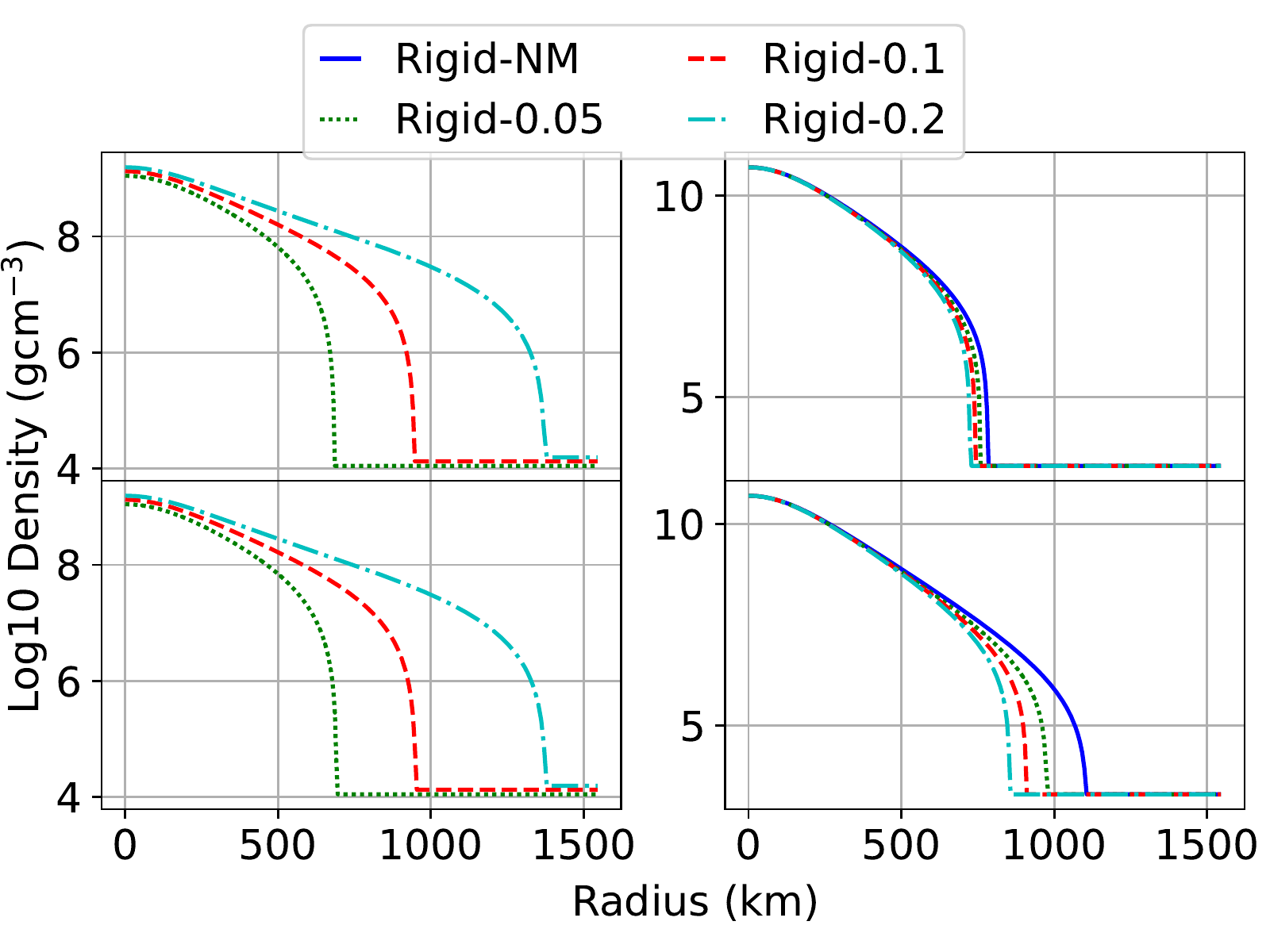}
	\caption{Initial density profiles for the rigidly-rotating DMAIC progenitors. The left (right) panel is for the DM (NM) component. The upper (lower) sub-panel in each panel is for the polar (equatorial) density profiles. \label{fig:dmnmrhoinitial}}
\end{figure}

\begin{figure}[ht!]
\centering
\includegraphics[width=1.0\linewidth]{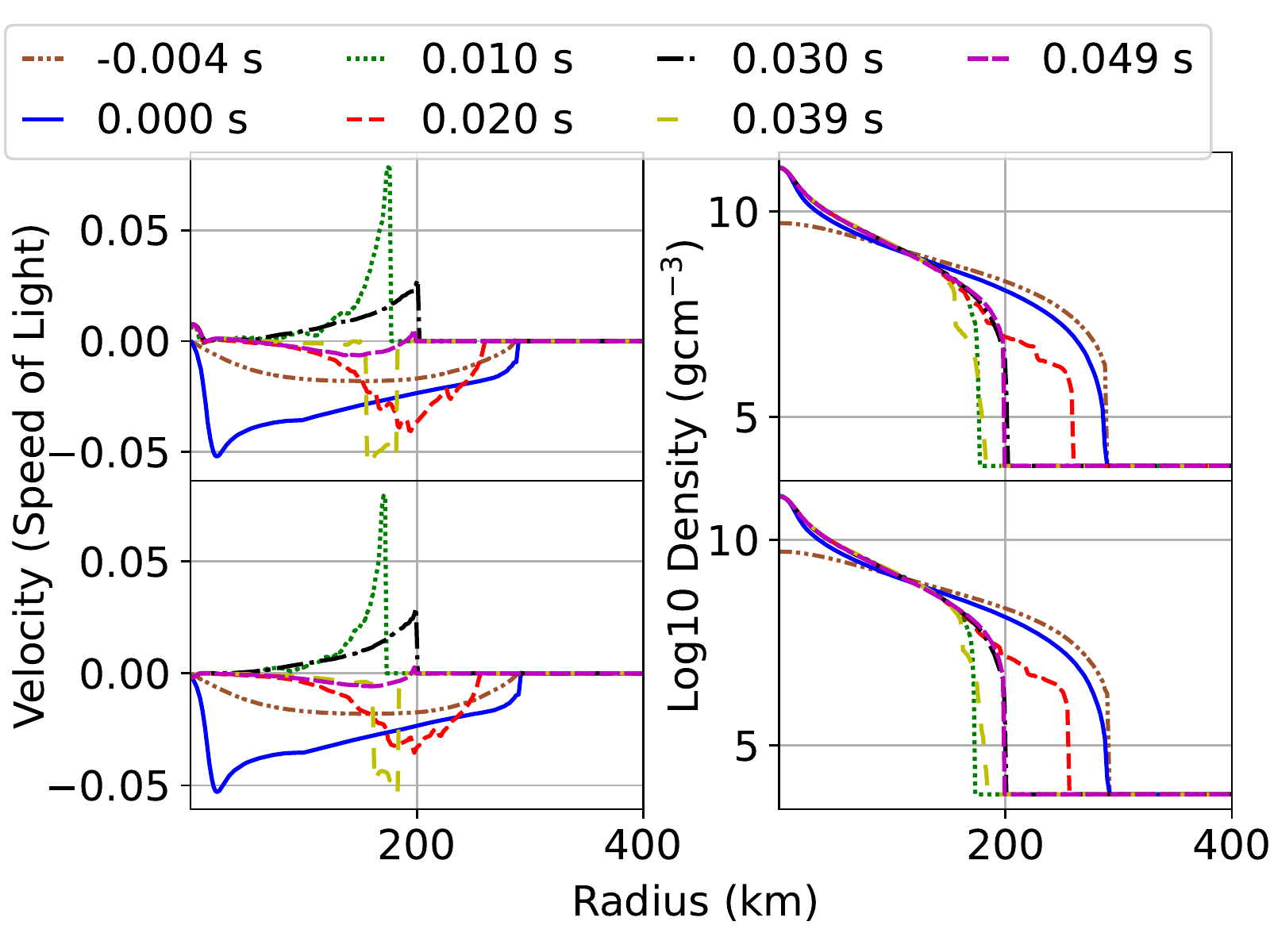}
\caption{Evolution of the DM radial density and velocity profiles of the Rigid-0.01 DMAIC model. The left (right) panel is for the velocity (density). The upper (lower) sub-panel in each panel is for the polar (equatorial) profiles. \label{fig:dmrhovel}}
\end{figure}

\begin{figure}[ht!]
\centering
\includegraphics[width=1.0\linewidth]{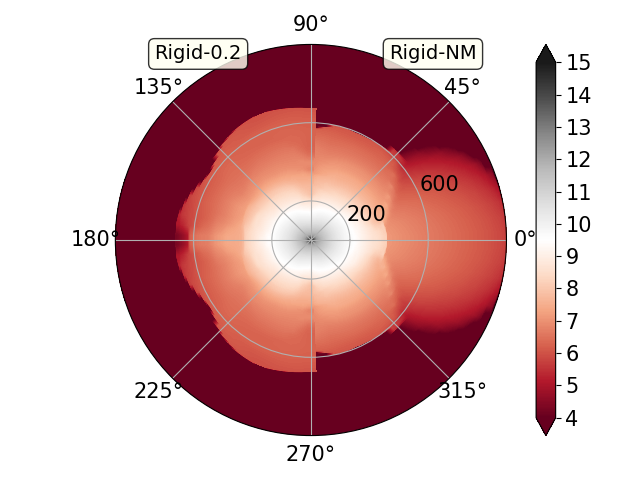}
\caption{NM density contour plot for two different rigidly-rotating DMAIC models at the end of the simulations. The right (left) plot is for the Rigid-NM (Rigid-0.2) model. Densities are in the log$_{10}$ scale of g cm$^{-3}$. The radial distance is in km. \label{fig:nmrhocontour}}
\end{figure}

We first focus on the collapse dynamics of DMAIC. From Table \ref{tab:tabmodels}, we observe that the admixture of DM delays the time of core bounce and reduces the proto-neutron star mass, which is similar to the results by \citet{2019ApJ...884....9L} and \citet{2019ApJ...883...13Z}. We show the maximum NM density evolution for the rigidly-rotating DMAIC models in the right panel of Figure \ref{fig:nmrhocentralrigid}. Despite having different initial and proto-neutron star masses (c.f. Table \ref{tab:tabmodels}), the maximum NM density evolution is almost identical for all DMAIC models. The final maximum NM densities are also insensitive to the DM mass fraction $\epsilon_{\text{DM}}$. Furthermore, we find that AIC is successful for all DMAIC progenitors. This differs from the results presented by \citet{2019ApJ...884....9L} and \citet{2019ApJ...883...13Z} because we assume different DM particle masses. Their work assumed a heavy ($1$ GeV) DM particle mass, leading to a more compact DM core with a large central density. Thus, it significantly impacts the NM density profile near its centre. The NM density decreases sharply due to the strong gravitational force provided by the compact DM core. Electron capture is less efficient in their model, so the NM component's effective adiabatic index remains near $\frac{4}{3}$. We assumed a light ($0.1$ GeV) DM particle mass in our study. The DM component is more diffusive and extended. Hence it brings a less significant impact to the NM density profile near its core. We show how the NM density profile changes with increasing DM mass fraction $\epsilon_{\text{DM}}$ in the right panel of Figure \ref{fig:dmnmrhoinitial}. We observe that when more DM is admixed, the core of the NM component remains almost unchanged. Since the collapse dynamics of a WD are governed by the dense core, where $\rho_{2}$ is large enough to initiate electron capture, it is natural to expect generic collapse dynamics for all rigidly-rotating DMRWDs. \\

We find that the DM component collapses with the NM component to form a bound DM core. We show the DM density profile evolution in the left panel of Figure \ref{fig:dmrhovel}. The DM density evolves similarly to the NM density, but it remains stable after the NM core bounce. We show the DM density profile evolution for a particular model Rigid-0.01 in the right panel of Figure \ref{fig:dmrhovel} as an example. The DM radius contracts from $\sim 350$ km at $\bar{t} = 0.014$ s, to $\sim 180$ km at $\bar{t} = 0.01$ s. Although the DM radius increases at $\bar{t} = 0.02$ s, the DM component gradually contracts to $\sim 200$ km at $\bar{t} = 0.03$ s and pulsates around $\sim 180 - 200$ km. This suggests that a bound DM component has formed with negligible mass loss. We show the DM velocity profile evolution of the same DMRWD model in the left panel of Figure \ref{fig:dmrhovel}. The post-bounce velocity shock breaks through the DM surface around $\bar{t} = 0.01$ s. However, the shock is too weak to unbind the DM component. The shock gradually weakens and becomes a sound wave that propagates inside the DM component. This also explains the pulsation of the DM component between $\bar{t} = 0.03$ and $0.049$ s.


\subsubsection{The Formation of DM-admixed Neutron Stars} \label{subsubsec:dmns}

What are the astrophysical implications of our findings? DM-admixed neutron stars have been extensively studied in the past decade. For instance, \citet{2020EPJC...80..544B} showed that the admixture of DM can explain the cooling rate of some pulsars/neutron stars, such as PSR B0656+14, PSR B1706-44 and PSR B2334+61, which could not be explained if the popular APR equation of state (EOS) is assumed. \citet{2021PhRvD.104f3028D} and \citet{2021ApJ...922..242L} discuss the anomalous $2.6$ $M_{\odot}$ object from the gravitational-wave event GW190814 \citep{2020ApJ...896L..44A} as a possible DM-admixed neutron star. However, the formation channel of DM-admixed neutron stars has never been addressed in depth. Although \citet{Zha2019} performed DMAIC simulations, their work assumed that the DM is compact and static. Our self-consistent, two-fluid simulations show that the AIC of a DMRWD would produce a DM-admixed (rotating) neutron star, such that the DM component is gravitationally bound with negligible mass loss. The collapse of DM also happens with a time scale similar to that of NM. Therefore, we have shown numerically that it is possible to form a DM-admixed neutron star through DMAIC.


\subsubsection{Gravitational-wave Signatures} \label{subsubsec:gwave}

\begin{figure}[ht!]
	\centering
	\includegraphics[width=1.0\linewidth]{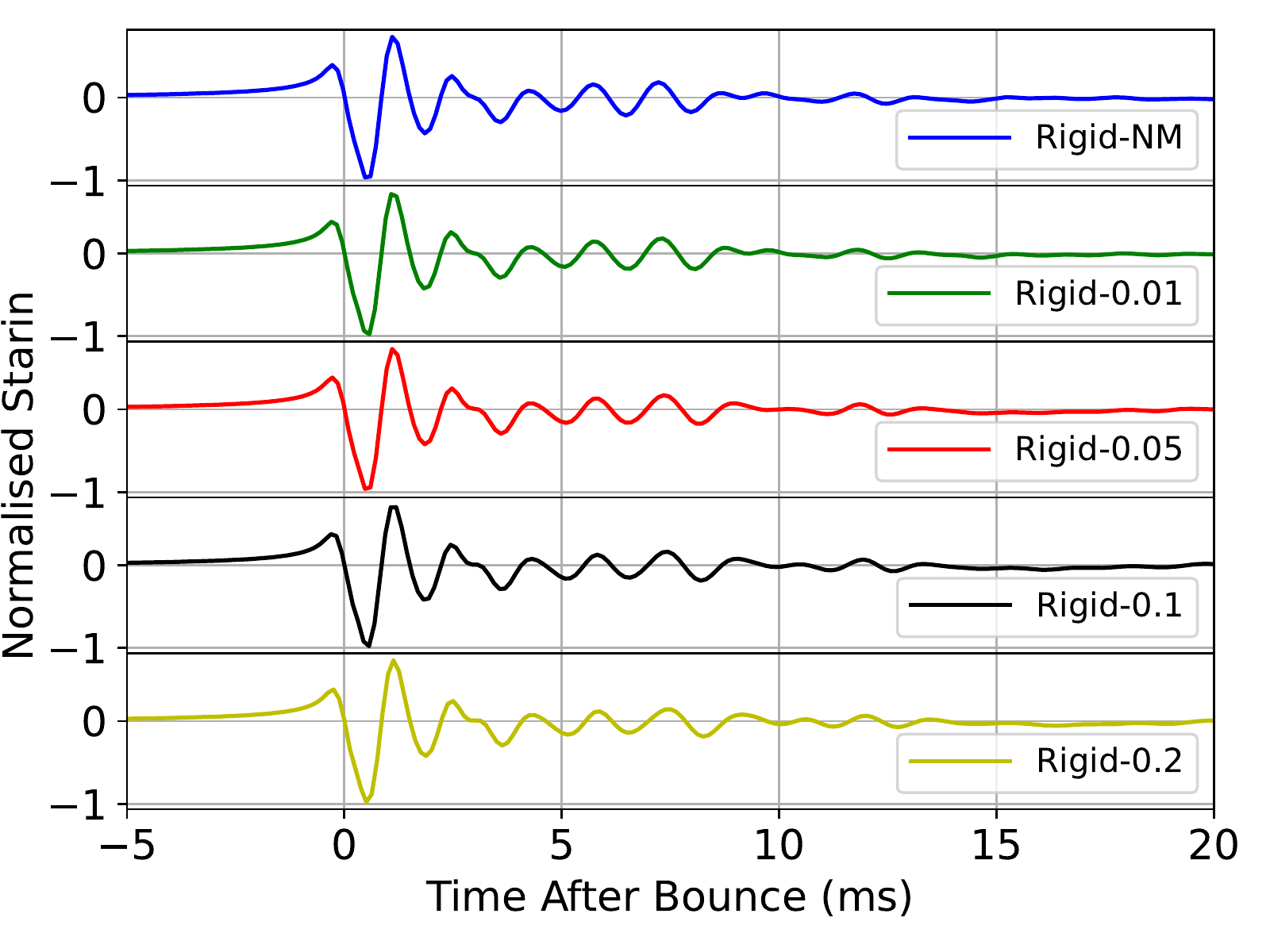}
	\caption{Total GW strains for the rigidly-rotating DMAIC models. We normalised all the GW strains to the corresponding maximum amplitude of the Rigid-NM model. The normalisation constant is $7.53\times10^{-21}$. \label{fig:gwaverigid}}
\end{figure}

\begin{figure}[ht!]
	\centering
	\includegraphics[width=1.0\linewidth]{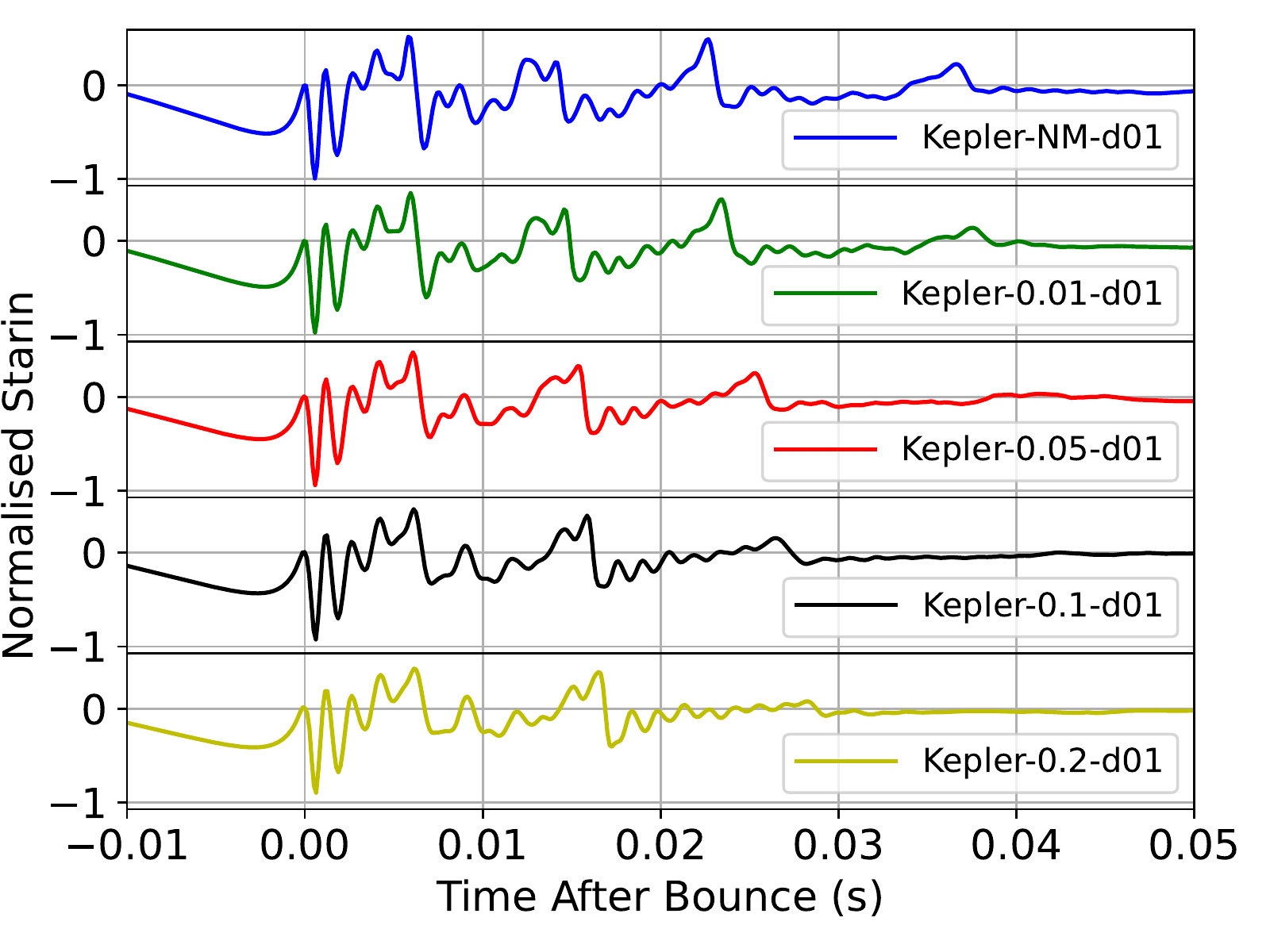}
	\caption{Same as Figure \ref{fig:gwaverigid}, but for the Kepler-rotating and $\alpha_{d} = 0.1$ models with a normalisation constant of $5.05\times10^{-21}$. \label{fig:gwavekeplerd01}}
\end{figure}

\begin{figure}[ht!]
	\centering
	\includegraphics[width=1.0\linewidth]{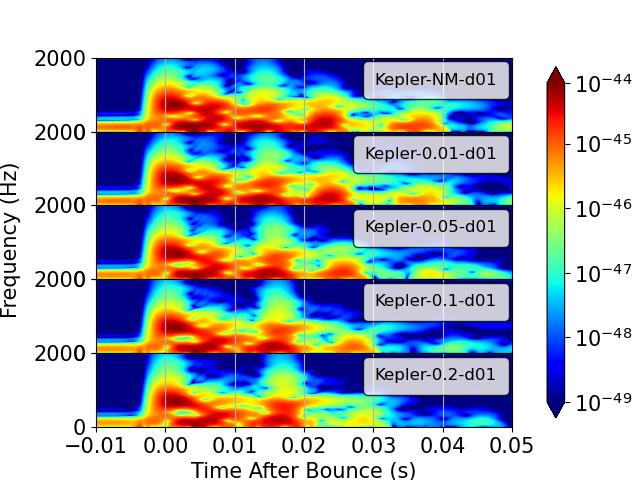}
	\caption{Power spectral density of DMAIC GWs for DMRWDs rotating in the Kepler rule with $\alpha_{d} = 0.1$.  \label{fig:spectrogramd01}}
\end{figure}

\begin{deluxetable}{lccc}
\caption{GW mismatch (in \%) with respect to the pure NM model for DMAICs with different initial rotation profiles. See Table \ref{tab:tabmodels} for the simulation parameters of these models. \label{tab:mismatch}}
\tablewidth{0pt}
\tablehead{
\colhead{-} & \colhead{Rigid} & \colhead{Kepler-d001} & \colhead{Kepler-d01}
}
\startdata
DM-0.01 & 0.306 & 13.570 & 19.217 \\
DM-0.03 & 0.870 & 26.151 & 42.386 \\
DM-0.05 & 1.263 & 29.103 & 41.173 \\
DM-0.07 & 1.596 & 36.149 & 41.169 \\
DM-0.09 & 1.844 & 38.256 & 44.517 \\
DM-0.1 & 1.992 & 40.916 & 43.785 \\
DM-0.2 & 2.784 & 41.845 & 52.794
\enddata
\end{deluxetable}

The non-luminous nature of the DM makes it difficult to be detected through conventional telescopes. The weak electromagnetic signatures from a typical AIC also hinder indirect DM detection by comparing AIC luminosities. Therefore, we rely on the GW signatures generated by both the NM and DM components. \\

Equation \ref{eqn:ecap} suggests that the moment of inertia tensor $I_{zz}$ is separable into individual DM and NM components:
\begin{equation}
\begin{aligned}
    I_{zz} = I_{zz,1} + I_{zz,2}, \\
    I_{zz,i} = \frac{1}{3}\int_{\text{All Space}}\rho_{i}r^{2}P_{2}(\text{cos}\theta) d\tau. \\
\end{aligned}
\end{equation}
Since the DM only interacts with NM through gravity, the Euler equation for the DM component does not contain any non-trivial NM-related terms except the gravitational potential $\Phi$. Hence, the GW signature from the AIC of a DMRWD can be separated into the DM and NM contributions:
\begin{equation}
\begin{aligned}
    h_{+} = h_{+,1} + h_{+,2}, \\
    h_{+,i} = \frac{3}{2}\frac{G}{Dc^{4}}\text{sin}^{2}\theta\frac{d^{2}}{dt^{2}}I_{zz,i}. \\
\end{aligned}
\end{equation}
To compute $h_{+,i}$, we make use of Equation (16) in \citet{2004ApJ...600..834O} and substitute all the components of $\vec{v}$ and $\rho$ by the corresponding DM/NM values. \\

It is also a common practice to study GW strains by time-frequency analysis. To obtain the GW spectrogram, we perform a windowed Fourier transform:
\begin{equation} \label{eqn:fourier}
    \tilde{h}^{*}(f, t) = \int_{-\infty}^{\infty}h_{+}(\tau)w(t, \tau)\text{exp}(-2\pi if\tau)d\tau.
\end{equation}
Here, $w(t, \tau)$ is the window function, and we choose the Hann window. \\

We first show the AIC GWs generated by the rigidly-rotating DMRWD models in Figure \ref{fig:gwaverigid}. The GWs are all generic Type-I waveforms \citep{2011LRR....14....1F}. There are no considerable differences in the GW signature with respect to all DM-admixed models. This contrasts with the results presented by \citet{Zha2019}, where they show enhanced amplitudes during $\bar{t} = 0$ s. This is because the contributions to the GW strains are mainly from the innermost core ($\sim 10$ km). We have shown in the previous section that the effects of admixing $0.1$ GeV DM on the NM density profile are mainly at the NM outer envelope. The NM collapse dynamics are also generic for all DM-admixed models. We append the NM density contour plots of models NM-Rigid and DM-Rigid-0.2 in Figure \ref{fig:nmrhocontour} for comparison. We observe that the dense core, which corresponds to the major part of the proto-neutron star of the DM-admixed model, is almost identical to that of the pure NM counterpart. This explains why the GW signatures from rigidly-rotating DMRWDs are all generic. \\

However, the situation is different for differentially rotating progenitors. We show the GW strains of the Kepler-rotating and $\alpha_{d} = 0.1$ model in Figure \ref{fig:gwavekeplerd01}. We find that the DM admixture indirectly suppresses the post-bounce 3rd and 4th peaks of the GW strains. This could also be observed as the gradual disappearance of the 3rd and 4th spectral peaks in Figure \ref{fig:spectrogramd01}. Therefore, the GW strains of DMAIC are qualitatively different from that of the pure NM model. We find that the spectral peaks exist for the pure NM model because the reflected shock waves pass through the NM core and make it pulsate non-radially. The corresponding pulsation amplitudes for the DM-admixed models are smaller, resulting in weaker GW signatures. We find similar results for the Kepler-rotating and $\alpha_{d} = 0.01$ model, except that the 4th spectral peak never exists for the pure NM and hence, the DM-admixed models. \\

The DM component is more diffusive for fermionic DM with a particle mass of $0.1$ GeV when compared to those with heavier DM particle mass. As such, the collapse dynamics of the DM component could not produce GW amplitudes comparable to that of the NM component. The effects of DM admixture on the total GW signatures are, therefore, indirect. To quantitatively determine whether such effects could be observable, we compute the mismatch $\mathfrak{M}$, which quantifies how dissimilar two waveforms are \citep{2011CQGra..28s5015R, 2017PhRvD..95f3019R}:
\begin{equation}
    \mathfrak{M} = 1 - \text{max}\left(\frac{\langle h_{a},h_{b} \rangle}{\sqrt{\langle h_{a},h_{a} \rangle\langle h_{b},h_{b} \rangle}}\right).
\end{equation}
The second term here contains the match between two waveforms $h_{a}$ and $h_{b}$:
\begin{equation} \label{eqn:inner}
    \langle h_{a},h_{b} \rangle = \int_{0}^{\infty} \frac{4\tilde{h_{a}}^{*}\tilde{h_{b}}}{s}df.
\end{equation}
Here, $s$ is the estimated noise amplitude spectral density of the Advanced LIGO \citep{ligopsd}. $\tilde{h}^{*}$ is the Fourier transform of the GW strain, which is just Equation \ref{eqn:fourier} but with $w(t, \tau) = 1$. The mismatch is maximized over the relative phase, amplitudes, and arrival times. We follow \citet{Zha2019} to set the integration limit of Equation \ref{eqn:inner} to be from 100 Hz to 2000 Hz. The computations are facilitated through the open-source package PyCBC \citep{alex_nitz_2022_6912865}. We extract GW waveforms for all the models listed in Table \ref{tab:tabmodels} with a time window of $-0.01$ s $ < \bar{t} < 0.05$ s and compute the mismatches with respect to the pure NM model. The results are listed in Table \ref{tab:mismatch}. The mismatches for the rigidly-rotating DMAIC models are small, which is no surprise because the GW waveforms of the DM-admixed models in such a scenario are very similar to that of the pure NM counterpart. The mismatches for the Kepler-rotating DMAIC models, however, are relatively large. The presence of a $1$\% of DM can be inferred from future GW detection produced by DMAIC if Advanced LIGO can distinguish two waveforms with an accuracy better than $14$\%.


\subsection{The Compact Dark Matter Limit} \label{subsec:compactdm}

\begin{deluxetable*}{ccccccccccccc}[ht!]
\tablecaption{Same as Table \ref{tab:tabmodels}, but for differentially rotating DMAIC progenitors that have $\alpha_{d} = 0.5$, $\Omega_{c} = 45.2 $ s$^{-1}$, and the DM particle mass of $0.3$ GeV. We also append the NM compactness $\mathfrak{C} = 2GM_{\rm NM}/R_{\rm eNM}c^{2}$. \label{tab:newmodels}}
\tablewidth{0pt}
\tablehead{
\colhead{Model} & \colhead{$M_{\rm NM}$} & \colhead{$M_{\rm DM}$} & \colhead{log$_{10}\rho_{1c}$} &
\colhead{$R_{\rm eNM}$} & \colhead{$R_{\rm eDM}$} & \colhead{$\epsilon_{\rm DM}$} & \colhead{$t_{b}$} & \colhead{log$_{10}\rho_{1b}$} & \colhead{log$_{10}\rho_{2b}$} & \colhead{$M_{\rm PNS}$} & \colhead{$\mathfrak{C}$} \\
\colhead{-} & \colhead{($M_{\odot}$)} & \colhead{($M_{\odot}$)} & \colhead{(gcm$^{-3}$)} &
\colhead{(km)} & \colhead{(km)} & \colhead{-} & \colhead{(ms)} & \colhead{(gcm$^{-3}$)} & \colhead{(gcm$^{-3}$)} & \colhead{($M_{\odot}$)} & \colhead{($10^{-3}$)}
}
\startdata
Kepler-NM-d05 & 1.771 & - & -  & 916 & - & 0.00 & 29.087 & - & 14.351 & 1.604 & 5.71 \\
Kepler-0.01-d05 & 1.672 & 0.017 & 9.968 & 929 & 157 & 0.01 & 30.595 & 12.720 & 14.345 & 1.516 & 5.32 \\
Kepler-0.03-d05 & 1.525 & 0.047 & 10.206 & 948 & 187 & 0.03 & 32.426 & 12.826 & 14.339 & 1.379 & 4.75 \\
Kepler-0.05-d05 & 1.410 & 0.074 & 10.313 & 961 & 202 & 0.05 & 33.778 & 12.856 & 14.336 & 1.267 & 4.33 \\
Kepler-0.07-d05 & 1.313 & 0.099 & 10.383 & 967 & 212 & 0.07 & 34.899 & 12.866 & 14.332 & 1.170 & 4.01 \\
Kepler-0.09-d05 & 1.229 & 0.122 & 10.434 & 967 & 220 & 0.09 & 35.888 & 12.871 & 14.332 & 1.090 & 3.75 \\
Kepler-0.1-d05 & 1.191 & 0.132 & 10.454 & 967 & 223 & 0.1 & 36.348 & 12.873 & 14.333 & 1.052 & 3.64 \\
Kepler-0.2-d05 & 0.896 & 0.224 & 10.588 & 935 & 246 & 0.2 & 40.396 & 12.872 & 14.331 & 0.767 & 2.83
\enddata
\end{deluxetable*}

The properties of a Fermionic DM-admixed compact star were shown to be sharply changing around DM particle mass of $0.1$ GeV \citep{PhysRevD.105.123010}. To better capture the transitional effects from a sub-GeV to GeV mass, we include progenitor models admixed with fermionic DM of particle mass $0.3$ GeV. Furthermore, the progenitors are all differentially-rotating DMRWDs with $\alpha_{d} = 0.5$. For reference, we include the parameters of our appended models in Table \ref{tab:newmodels}. We generally find similar collapse dynamics for the DM and NM components as those of the diffusive DM limit. For instance, we find a delay in the NM bounce time and the successful formation of a DM-admixed neutron star. The in-depth discussion of the collapse dynamics of DMAIC under the compact DM limit would therefore be omitted.


\subsubsection{Gravitational Waves from the Dark Matter Component} \label{subsubsec:dmgwave}

\begin{figure}[ht!]
	\centering
	\includegraphics[width=1.0\linewidth]{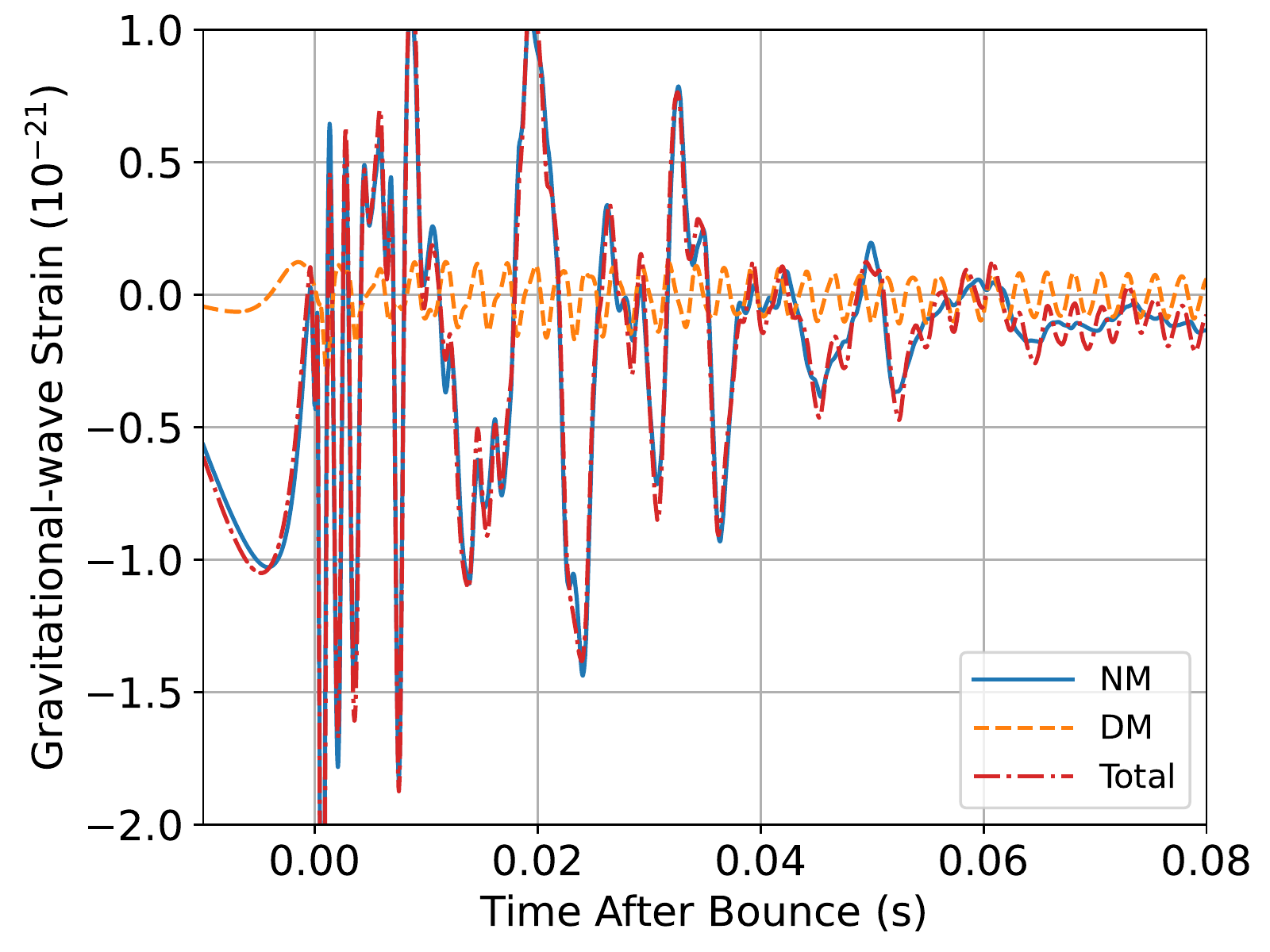}
	\caption{Magnified plot of the NM (blue solid line), DM (orange dashed line) and, total (red dashed-dotted line) GW strains for the Kepler-0.05-d05 model. Here, a distance of $D = 10$ kpc to the AIC is assumed.  \label{fig:dmgwaveall}}
\end{figure}

\begin{figure}[ht!]
	\centering
	\includegraphics[width=1.0\linewidth]{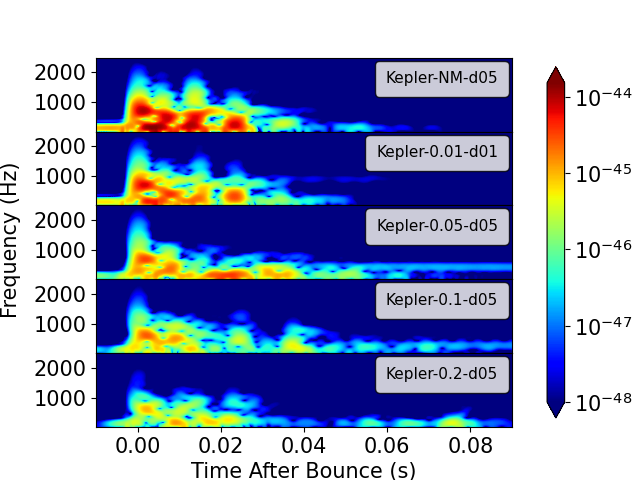}
	\caption{Same as Figure \ref{fig:spectrogramd01}, but for DMRWDs rotating in the Kepler rule with $\alpha_{d} = 0.5$.  \label{fig:spectrogramd05}}
\end{figure}

\begin{figure}[ht!]
	\centering
	\includegraphics[width=1.0\linewidth]{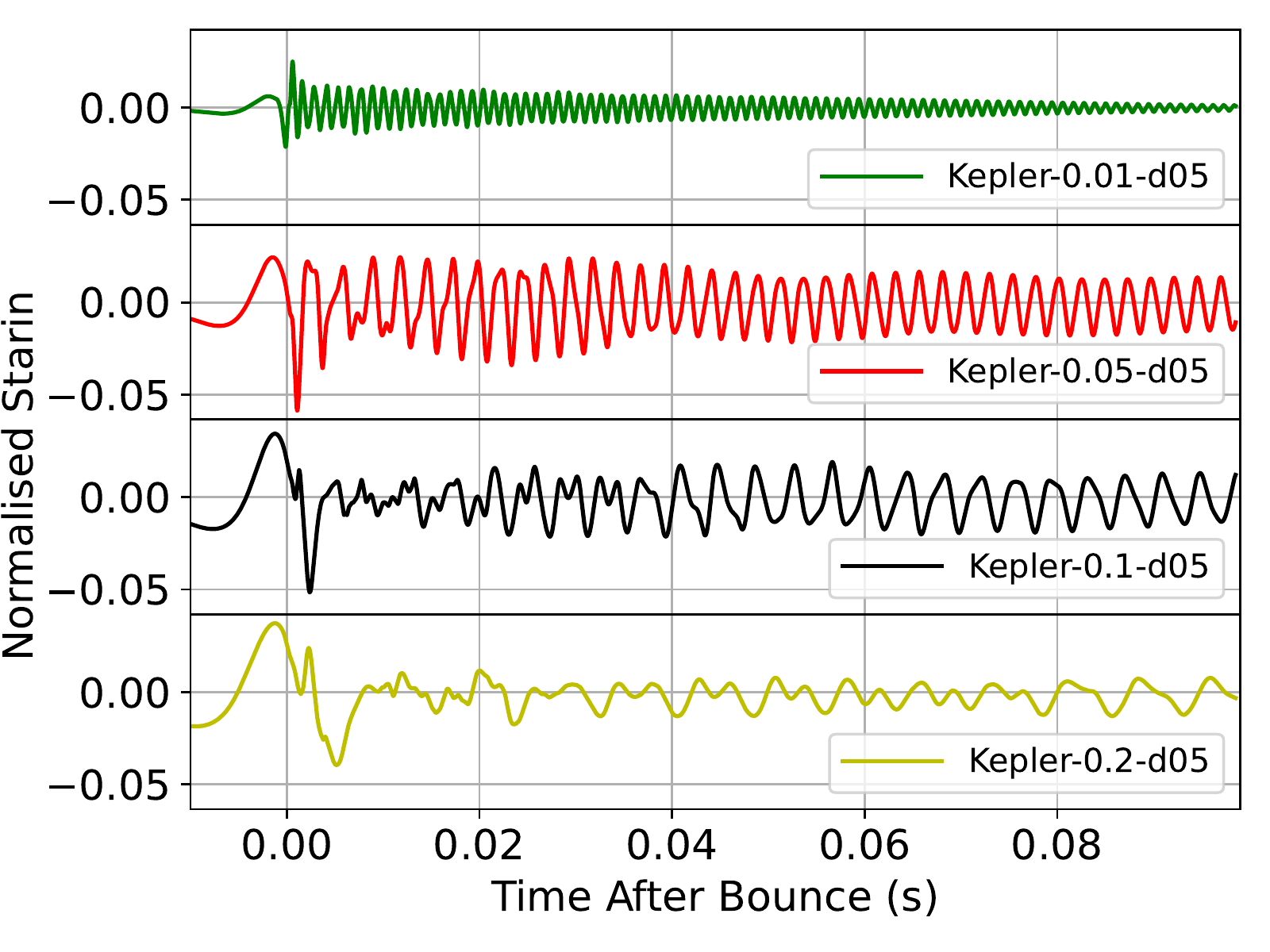}
	\caption{Same as Figure \ref{fig:gwaverigid}, but for the DM GWs of the Kepler-rotating DMAIC models with $\alpha_{d} = 0.5$ only. The normalisation constant is $6.07 \times 10^{-21}$.  \label{fig:dmonly}}
\end{figure}

In this section, we focus on the GW signature produced by a DMAIC event. We show the GW strains of a particular model Kepler-0.05-d05 in Figure \ref{fig:dmgwaveall}. The DM GW strain has amplitudes comparable to that of the NM and produces secondary oscillations on top of the NM GW strain. This is in contrast to the diffusive DM limit. The DM component, which couples to a highly differentially rotating NM configuration set by $\alpha_d = 0.5$, becomes spheroidal in shape. The vigorous non-spherical collapse dynamics thus generate a considerable magnitude of GWs. We analyze the GW strains by plotting their spectrograms in Figure \ref{fig:spectrogramd05}. We find that the admixture of DM greatly suppresses the NM GW strains around the NM core bounce, which is no surprise because the NM mass and compactness are reduced substantially when DM is admixed (c.f. Table \ref{tab:newmodels}). We also find that the DM GW strains show up as a continuous low-frequency ($< 1000$ Hz) signal in the spectrogram before $\bar{t} = 0.1$ s. This is important because convection-related GW signals are emitted after $\bar{t} = 0.1 - 0.2$ s \citep{alma991040185904203407}. Therefore, any low-frequency signals observed before $\bar{t} = 0.1$ s could be direct evidence of a compact DM admixture. The DM GW waveforms (see Figure \ref{fig:dmonly}) are consistent with the Type III collapsing polytrope waveforms presented in \citet{2011LRR....14....1F}.

\subsubsection{Detection Prospect} \label{subsubsec:detect}

\begin{figure}[ht!]
	\centering
	\includegraphics[width=1.0\linewidth]{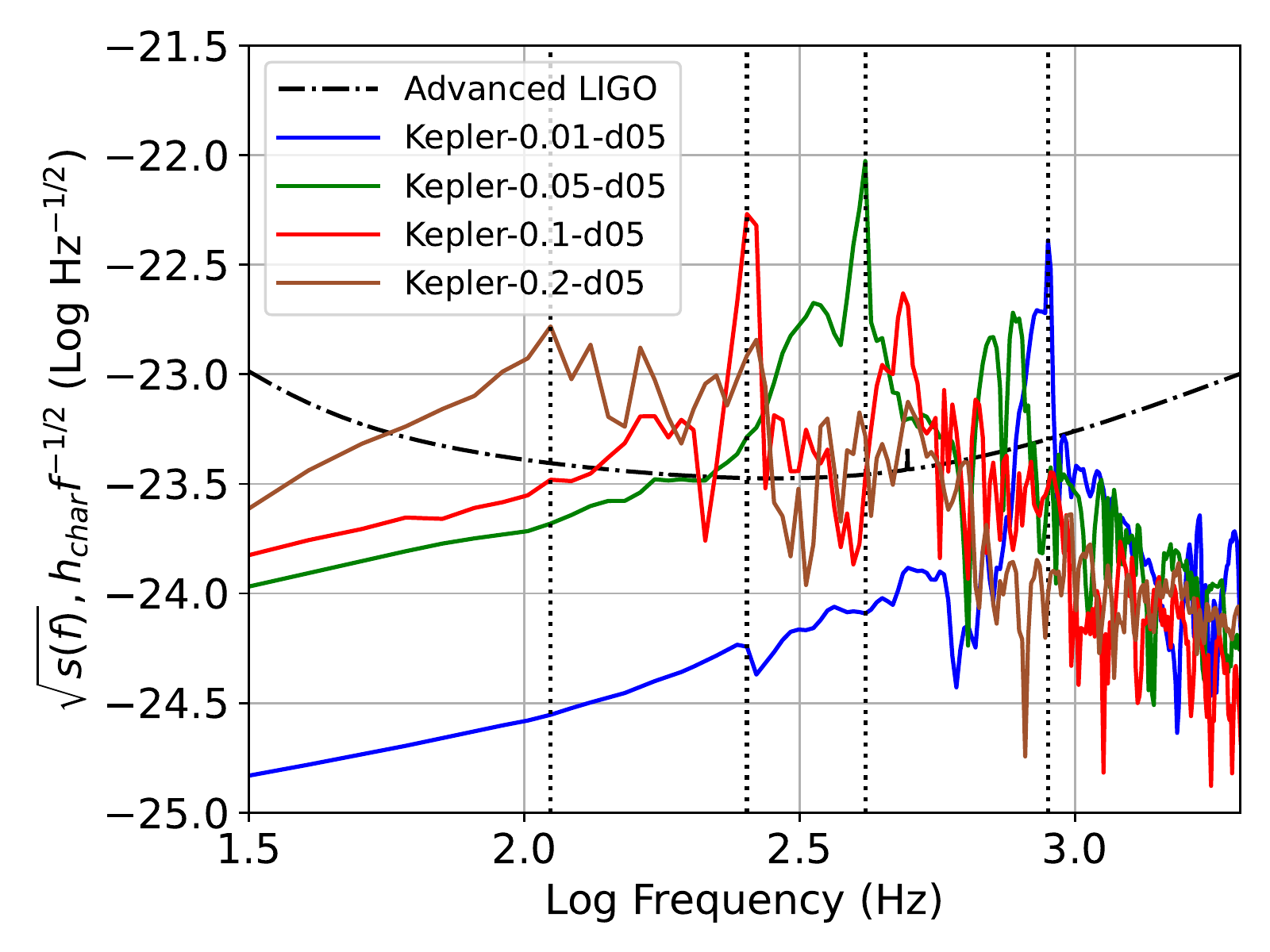}
	\caption{Scaled characteristic DM GW strains for the differentially-rotating DMAIC models with $\alpha_{d} = 0.5$. Peak frequencies obtained from the Fourier transform are marked as vertical black dotted lines. \label{fig:ligo}}
\end{figure}

\begin{figure}[ht!]
	\centering
	\includegraphics[width=1.0\linewidth]{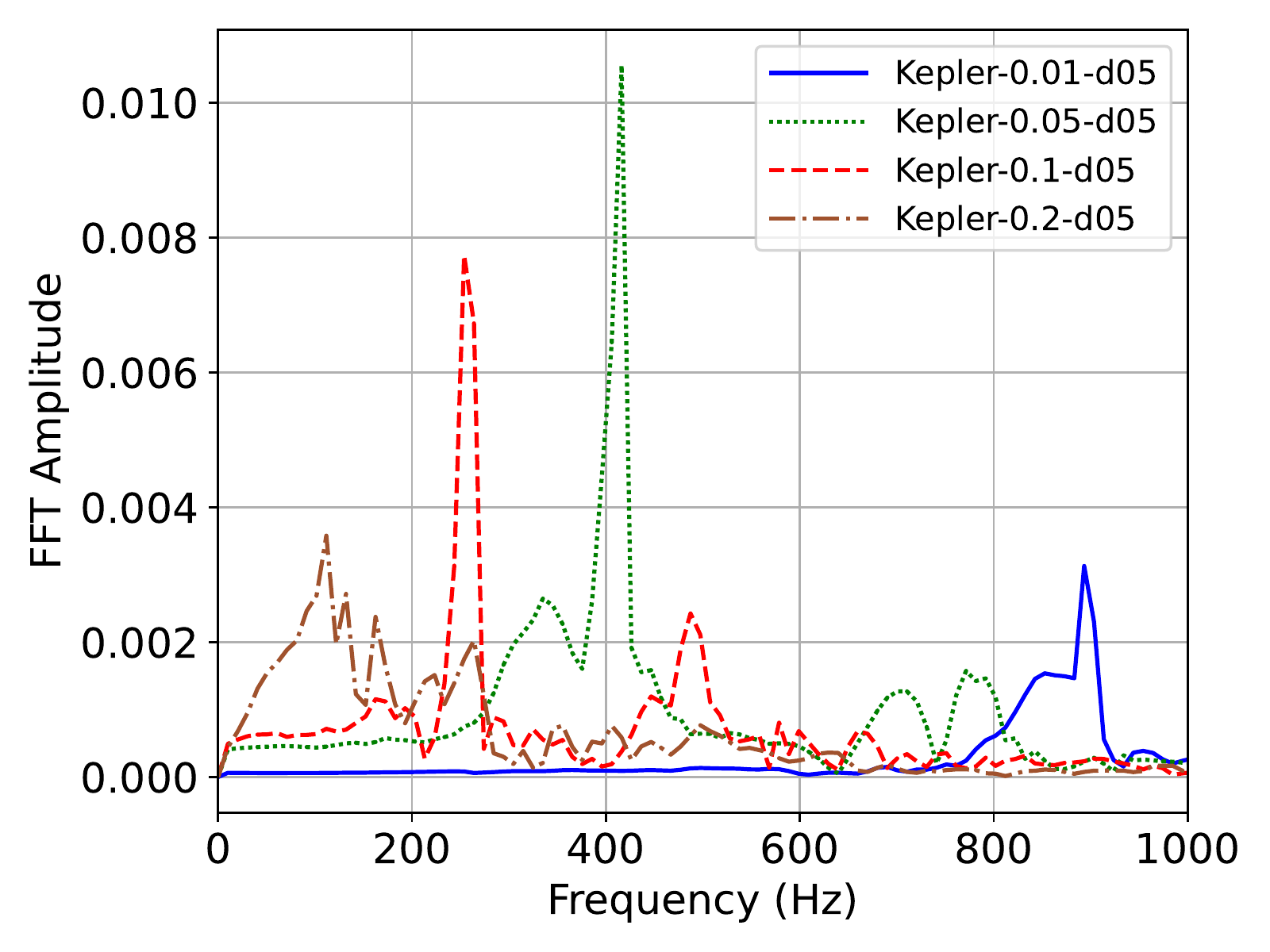}
	\caption{Fourier transformed amplitude of the DM GWs against frequency for 4 different DMAIC models with $\alpha_{d} = 0.5$. \label{fig:fft}}
\end{figure}

To study the detectability of the DM GW signals, we compute the dimensionless characteristic GW strain \citep{PhysRevD.57.4535}:
\begin{equation}
    h_{\text{char}} = \sqrt{\frac{2}{\pi^{2}}\frac{G}{c^{3}}\frac{1}{D^{2}}\frac{dE_{\text{GW}}}{df}}.
\end{equation}
Here, $\frac{dE_{\text{GW}}}{df}$ is the GW spectral energy \citep{2009ApJ...707.1173M}:
\begin{equation}
    \frac{dE_{\text{GW}}}{df} = \frac{3}{5}\frac{G}{c^{5}}(2\pi f)^{2}|\tilde{h}_{+}|^{2}.
\end{equation}
We compare $h_{\text{char}}f^{-1/2}$ with the Advanced LIGO noise spectral density $\sqrt{s(f)}$ in Figure \ref{fig:ligo}. In the same figure, we mark vertical lines corresponding to the peak frequencies of the DM GW waveforms (see Figure \ref{fig:fft}). We choose the sampling window as $\bar{t} > 0$ s. The DM characteristic GW strains corresponding to the frequency peaks are above the Advanced LIGO sensitivity curve, which is true for all of our considered models for all our considered models, assuming $D$ = 10 kpc. Hence, the GW signature of a collapsing, compact DM in a Milky Way DMAIC event should be detectable by Advanced LIGO. Our results represent the first-ever numerical calculation of the GW waveforms of a collapsing DM core in a compact star. Finally, we show the detectability of DM GWs from rigidly rotating progenitors in Appendix \ref{sec:rigidwave}.


\section{Conclusion} \label{sec:conclusion}

We presented two-dimensional simulations of DMAIC with self-consistent modelling of the DM dynamics. Regardless of the DM particle mass and compactness, the DM component follows the collapse of the NM component to become a bound DM core with a time scale comparable to that of the NM. This result demonstrates numerically, for the first time, how a DM-admixed neutron star could form through DMAIC. We also find that the NM bounce time is delayed, and the proto-neutron star mass is reduced when DM is admixed, similar as found in \citet{2019ApJ...884....9L} and \citet{2019ApJ...883...13Z}, where the DM component is modelled as a fixed compact core. \\

Due to the weak electromagnetic signals produced by the gravitational collapse of WDs, GW becomes an important and reliable channel to detect and study AIC. We computed the GW signatures for the NM and DM components using the quadrupole formula. For DM with a particle mass of $0.1$ GeV, the DM component is more diffusive and extended. Hence, the collapse of the DM component does not produce a significant GW signal. However, the admixture of such DM indirectly influences the NM signal by suppressing the NM GW spectral peaks after the NM core bounce. The significant alteration of the NM GW frequency spectrum also makes the DMAIC waveforms easily detectable by GW detectors, which show a $14$\% mismatch with the pure NM counterpart with only $1$\% of DM admixed. For DM with a particle mass of $0.3$ GeV, the DM component is more compact when compared to those with a particle mass of $0.1$ GeV. The admixture of DM greatly reduces the NM mass and hence its compactness. The NM GW signal at bounce is therefore decreased substantially. However, the DM component is massive and compact enough to produce a GW signal comparable to that of the NM counterpart during its dynamical collapse. The DM GW add to the NM GW and show up as secondary oscillations. These oscillations could be seen as continuous low-frequency ($< 1000$ Hz) signals in the GW spectrogram, occurring at $\bar{t} < 0.1$ s, which is before the time of low-frequency GW induced by prompt convection, providing direct evidence of the existence of DM. All the peak-frequency signals of the DM component in our models of a Milky Way DMAIC event are detectable by the Advanced LIGO. Our result is the first-ever computation of GW from a collapsing DM core, and these findings could provide the key features to identify DM in AIC events through future GW detections. \\

There are possible future improvements to our calculations. First, we assumed the DM component to be non-rotating, which could be relaxed to allow the DM to have collective motion, such as rotation, should there be adequate self-interaction of the DM. Second, we omitted detailed neutrino-transport physics in the simulations. Whether the presence of DM would significantly affect neutrino-flavour production would be an interesting future study. Lastly, we only include ad-hoc relativistic corrections to the gravity and dynamical equations. A more accurate picture of the collapse dynamics and the GW signature would call for solving the dynamical equations in the full general relativistic framework.


\acknowledgments
We thank Otto Akseli Hannuksela for his helpful discussion regarding gravitational-wave mismatch calculations. This work is partially supported by a grant from the Research Grant Council of the Hong Kong Special Administrative Region, China (Projects No. 14300320 and 14304322). Shing-Chi Leung acknowledges support from NASA grants HST-AR-15021.001-A and 80NSSC18K1017.


\appendix

\begin{figure*}[ht!]
\gridline{\fig{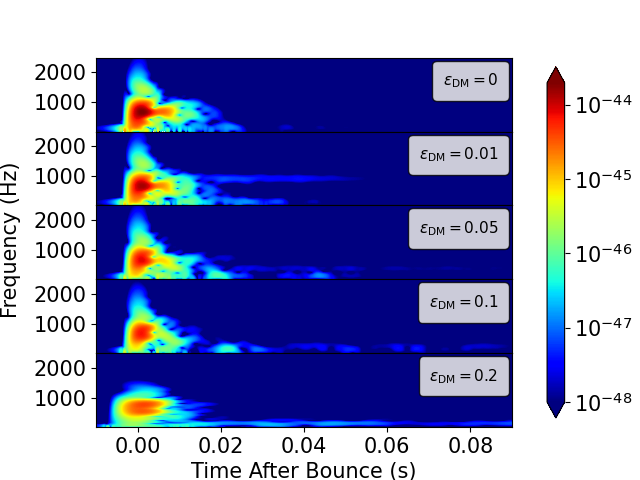}{0.5\textwidth}{(a)}
          \fig{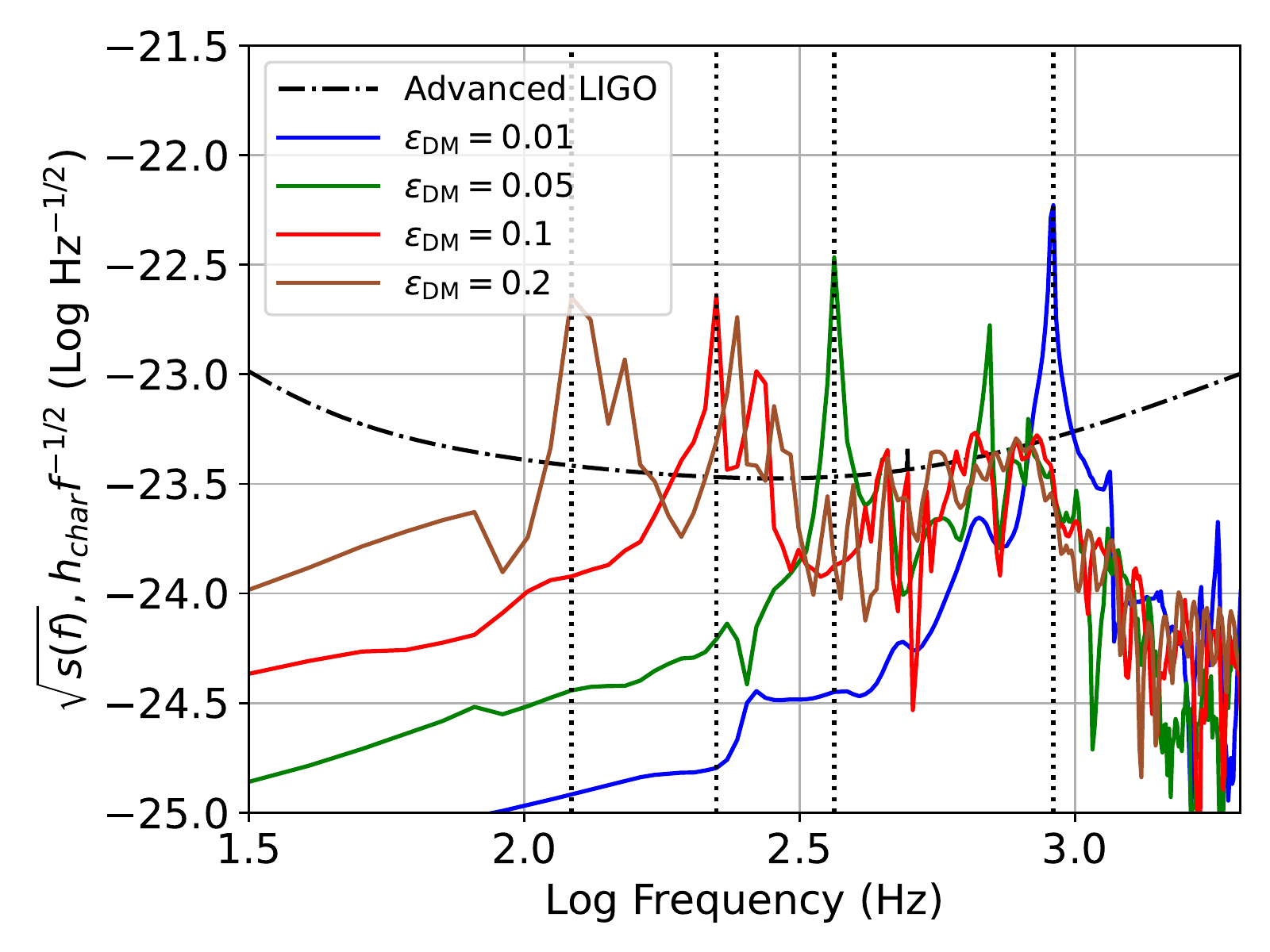}{0.5\textwidth}{(b)}
         }
\caption{(a) Power spectral density of DMAIC GWs for rigidly-rotating progenitors with increasing DM mass fraction $\epsilon_{\text{DM}}$. (b) Same as Figure \ref{fig:ligo}, but for characteristic wave strains of models presented in (a) and their comparison with the Advanced LIGO sensitivity curve (dashed line). \label{fig:rigid-appendix}}
\end{figure*}

\section{Formation of Dark Matter-admixed White Dwarf} \label{sec:dmawdcreate}

We follow \citet{Chan_2022} to consider the progenitor of DMRWD to be a star born with an inherent admixture of DM. We assume the DM and NM to be spherically symmetric clouds having constant densities $\rho_{1}$ and $\rho_{2}$, respectively. We consider the situation with the DM radius $R_{1}$ being larger than that of the NM, $R_{2}$. The total energy $E$ is:
\begin{equation}
\begin{aligned}
	E = -\left(\frac{3}{5}\frac{GM_{1}^{2}}{R_{1}} + \frac{3}{5}\frac{GM_{2}^{2}}{R_{2}} + \frac{3}{2}\frac{GM_{1}M_{2}}{R_{1}} - \frac{3}{10}\frac{GM_{1}^{2}R_{1}^{2}}{R_{1}^{3}}\right)
	\\+ \frac{3}{2}NkT + \frac{1}{2}M_{1}v_{1}^{2}.
\end{aligned}
\end{equation}
Here, $v_{1}$ is the DM ``thermal'' velocity, $N = M_{2}/m_{\text{H}}$ is the total number of NM nuclei, and $m_{H}$ is the molecular mass of hydrogen. Furthermore, we assume an extreme case of $M_{1} \sim 0.1$ $M_{\odot}$, $M_{2} \sim 10.0$ $M_{\odot}$. For a typical collapsing molecular cloud, we have $T \sim 150$ K and $\rho_{2} \sim 10^{8}m_{\text{H}}$ cm$^{-3}$, and hence $R_{2} = 3.05\times 10^{16}$ cm is smaller than the Jeans radius. We solve $E(R_{2})$ = 0 to obtain the maximum DM velocity of $v_{1\text{max}} \sim 1.27\times 10^{6}$ cm s$^{-1}$. Any $v_{1} < v_{1\text{max}}$ would give us a set of solution for $R_{1}$ and $\rho_{1}$. However, the most probable DM speed (assuming a Maxwell distribution) is $v_{\text{p}1} \sim 10^{7}$ cm s$^{-1}$. To take the velocity of DM into account, the bounded DM fraction is given by $f$:
\begin{align}
	f = \frac{\int_{0}^{u_{1}}u^{2}\text{exp}(-u^{2})du}{\int_{0}^{\infty}u^{2}\text{exp}(-u^{2})du}.
\end{align}
Here, $u = v/v_{\text{p}1}$, and $u_{1} = v_{1}/v_{\text{p}1}$. We take a particular $v_{1} = 1.23\times 10^{6}$ cm s$^{-1}$, and give two sets of solutions in $(R_{1}, \rho_{1})$ for $E < 0$: ($1.71\times10^{18}$ cm, 3860 GeV/cm$^{3}$) and ($6.10\times10^{16}$ cm, $8.48\times10^{7}$ GeV/cm$^{3}$). The required DM density in the first set of solutions is based on the state-of-the-art simulations, which showed that the DM density at the galactic bulge could be $\sim 3600$ GeV cm$^{-3}$ \citep{2014MNRAS.445.3133P}. The required DM density in the other set of solutions is much larger. However, such a value is possible near the galactic centre, and values with a similar order of magnitude have been adopted in studying the effect of DM annihilation on main-sequence stars \citep{2006astro.ph..8535M, 2008ApJ...677L...1I}. In conclusion, our estimations considering the DM velocity dispersions show that it is possible to trap a DM of $0.1$ $M_{\odot}$ during the star-forming phase, provided that the molecular cloud is in the vicinity of the galactic centre. There might be concern about whether the DM would follow the collapse of the NM to form a composite bound object. We show in an earlier section that a collapsing NM component would eventually induce a collapsing DM component to form a DM-admixed stellar object. This would, in our case, be a DM-admixed neutron star. Also, the collapse of the DM component happens with a time scale comparable to that of the NM, regardless of its size and mass. By simple scaling FFOR EXArelations, we can qualitatively conclude that the same scenario should also hold for molecular cloud collapse. Therefore, a zero-age main sequence with an inherent DM admixture should be possible, though a detailed numerical simulation shall be employed to justify our conjecture.

\section{Dark Matter Gravitational Waves for Rigidly-rotating Progenitors } \label{sec:rigidwave}

In section \ref{subsubsec:detect} we show the features of DM GWs from differentially-rotating progenitors and demonstrate that they are detectable by the Advanced LIGO, provided that the DM particle mass is $0.3$ GeV. Here, we perform a similar analysis for rigidly-rotating progenitors. In Figure \ref{fig:rigid-appendix} (a), we show the GW spectrograms for rigidly-rotating progenitors. These progenitors are rotating at $\sim 0.97$ that of the critical velocity and have increasing DM mass fraction $\epsilon_{\text{DM}}$ from $0$ to $0.2$. We observe that the DM GWs can also be captured as continuous low-frequency ($< 1000$ Hz) signals. In Figure \ref{fig:rigid-appendix} (b), we observe that all peak frequency signals of the DM GWs are detectable by the Advanced LIGO. 


\bibliography{main}{}
\bibliographystyle{aasjournal}

\end{document}